\documentclass[runningheads]{llncs}
\usepackage[T1]{fontenc}
\usepackage{graphicx}
\usepackage{marginnote}
\newcommand{\pageenlarge}[1]{\marginnote{}\enlargethispage{#1\baselineskip}}

\usepackage{booktabs} 
\usepackage{multirow}
\usepackage{makecell}
\usepackage{graphicx}
\usepackage{array}    
\usepackage{subcaption}
\usepackage{amsmath}
\usepackage{xcolor}
\usepackage{orcidlink}

\begin{document}
\title{Who Benefits from RAG? The Role of Exposure, Utility and Attribution Bias}
\titlerunning{The Role of Exposure, Utility and Attribution Bias in RAG}
%
    \author{Mahdi Dehghan\orcidlink{0000-0001-8843-4652} \and Graham McDonald\orcidlink{0000-0002-1266-5996}}
\institute{University of Glasgow, Glasgow, UK \\
\email{m.dehghan.1@research.gla.ac.uk}\\
\email{graham.mcdonald@glasgow.ac.uk}
}

\maketitle              

\begin{abstract}
\looseness=-1 Large Language Models (LLMs) enhanced with Retrieval-Augmented Generation (RAG) have achieved substantial improvements in accuracy by grounding their responses in external documents that are relevant to the user's query. However, relatively little work has investigated the impact of RAG in terms of fairness. Particularly, it is not yet known if queries that are associated with certain groups within a fairness category systematically receive higher accuracy, or accuracy improvements in RAG systems compared to LLM-only, a phenomenon we refer to as query group fairness. In this work, we conduct extensive experiments to investigate the impact of three key factors on query group fairness in RAG, namely: Group exposure, i.e., the proportion of documents from each group appearing in the retrieved set, determined by the retriever; Group utility, i.e., the degree to which documents from each group contribute to improving answer accuracy, capturing retriever–generator interactions; and Group attribution, i.e., the extent to which the generator relies on documents from each group when producing responses. We examine group-level average accuracy and accuracy improvements disparities across four fairness categories using three datasets derived from the TREC 2022 Fair Ranking Track for two tasks: \textit{article generation} and \textit{title generation}. Our findings show that RAG systems suffer from the query group fairness problem and amplify disparities in terms of average accuracy across queries from different groups, compared to an LLM-only setting. Moreover, group utility, exposure, and attribution can exhibit strong positive or negative correlations with average accuracy or accuracy improvements of queries from that group, highlighting their important role in fair RAG. Our data and code are publicly available from Github.\footnote{https://github.com/dehghanm/QueryGroupFairness\_in\_RAG/tree/main}

\end{abstract}


\section{Introduction} \label{introduction}
Retrieval-Augmented Generation (RAG) can improve the accuracy of generated responses from a Large Language Model (LLM) by supplementing the LLM with relevant documents that are retrieved from an external corpus. A typical RAG system comprises of two components: a retriever that retrieves documents that are relevant to the user's query, and a generator that produces answers conditioned on the retrieved documents. While many studies have examined the effectiveness of RAG in terms of accuracy, little attention has been paid in the literature to beyond-accuracy aspects, such as fairness~\cite{10.1145/3731120.3744599}. 

\looseness=-1 An underexplored aspect of fairness in RAG systems is \textit{query group} fairness. Query group fairness concerns whether a RAG system is systematically more accurate for queries that relate to particular groups within a fairness category, or when the inclusion of the retriever component in RAG results in greater accuracy improvements for particular query groups. A fairness category defines a dimension along which content can be grouped, while fairness groups correspond to the specific values within that fairness category. For example, consider a \textit{popularity} fairness category (with popular and unpopular groups). If a user queries the RAG system about an unpopular group, e.g., rural counties in the query ``Write an article on population trends in rural counties of the Great Plains, USA'', the RAG system may emphasise urban trends (i.e., the popular group) when generating the response, since most demographic studies focus on places with larger populations. In other words, since fewer demographic studies are about rural counties, compared to urban areas, the RAG system can be affected by either of the following two biases: (i) the generator may prioritise studies on larger cities when consuming the retrieved documents, or (ii) the rural counties may be unfairly underexposed to the generator due to fewer relevant documents in the retrieval results. In this scenario, the RAG system might systematically be less accurate for queries from the unpopular group. Moreover, the accuracy improvements, defined as the difference in the effectiveness of a RAG system compared to the LLM alone, could be smaller for a query from the unpopular group.

\pageenlarge 1
In this paper, we argue that three key factors are likely to influence query group fairness in RAG systems, namely: group exposure, i.e., the prevalence of documents from each group in the retrieved set, determined by the retriever; group utility, i.e., the extent to which documents from each group improve answer accuracy, reflecting retriever-generator interactions; and group attribution, i.e., the degree to which a generator consumes documents from each group when producing responses. To investigate their role in query group fairness, we build three datasets from the TREC 2022 Fair Ranking Track~\cite{Ekstrand_2022_fair_ranking_overview} test collection, where each dataset contains documents about a specific topic (i.e., Cities, Geography and Military History) and has ground truth labels for four fairness categories, each of which have four fairness groups. We define two tasks, namely article generation and title generation. Using our datasets, we first show the existence of the query group fairness issue in RAG systems, then examine how utility, exposure, and attribution scores of a group within a fairness category affect the accuracy or accuracy improvements of queries from that group.

\section{Related Works} \label{related_work}
The concept of RAG was first introduced by Lewis et al.~\cite{lewis2020retrieval}, and has been widely applied in various domains such as text summarization~\cite{edge2024local} and keyphrase suggestion~\cite{keyphrase}. Prior studies addressed both the effectiveness~\cite{jiang2023active,shao-etal-2023-enhancing} and efficiency~\cite{guo2025enhancing,li-etal-2023-compressing} of RAG systems. However, relatively little attention has been paid to assessing the performance of RAG systems from a fairness perspective~\cite{10.1145/3731120.3744599}. Fairness has been well examined in traditional information retrieval systems~\cite{fang2024fairness}, typically categorised into individual fairness, which ensures equal treatment of items within rankings, and group fairness, which concerns the representation of different item groups~\cite{zehlike2021fairness}. In RAG systems, fairness concerns extend beyond retrieval to how LLMs use retrieved documents in answer generation. 

Abdolghasemi et al.~\cite{abolghasemi2025evaluation} studied group-level fairness in answer attribution (i.e., attributing responses to the sources of information in the retrieved set) and showed that explicitly mentioning document's author (human vs. AI-generated) in the retrieved set can bias LLMs toward human-authored documents or assist them with the task of self-attribution. Building on this, we examine whether LLMs favor documents from specific groups within a fairness category when generating responses, and how such potential bias influences output accuracy. Wu et al.~\cite{wu2025does} introduced a scenario-based QA dataset derived from the TREC 2022 Fair Ranking Track~\cite{Ekstrand_2022_fair_ranking_overview} to examine group accuracy disparities in RAG. The authors tasked an LLM with choosing the most relevant answer to a topic from two options which may belong to a protected or unprotected group. They evaluated fairness by measuring how often the RAG system favored one group over the other. Their findings showed that RAG systems often amplify unfairness, though they generally improve response accuracy compared to a LLM-alone. Inspired by~\cite{wu2025does}, we use the TREC 2022 Fair Ranking Track to construct datasets for investigating query group fairness in RAG systems.

\pageenlarge 1
Salemi et al.~\cite{salemi2024evaluating} introduced eRAG evaluation framework for assessing retriever quality in RAG systems. In their work, each retrieved document, along with the query, was individually fed to the LLM, and its usefulness was defined by the correctness of the generated response with respect to ground truth value. Ranking quality was then derived from document usefulness. Building on eRAG, Kim et al.~\cite{kim_fairrag} defined document utility as whether asking the LLM a query along with a single document improves response correctness compared to prompting the LLM with query alone. Then, they examined the role of a stochastic re-ranker which increases individual document's exposure by re-ordering based on documents' utilities, in ensuring fair consumption of individual documents in generated responses. To determine document consumption, they applied Natural Language Inference based answer attribution~\cite{gao2023rarr,gao2023enabling,honovich2022true,muller2023evaluating}, which assess if a retrieved document entails the generated response. Building upon their idea, we extend the analysis from individual level to group level. Particularly, we employ NLI-based answer attribution, together with measures of document utility and exposure, to investigate how these factors, considered at the group level, impact the accuracy of generated responses across queries from different groups.

\section{Problem Statement} \label{problem_statement}
Let $\mathcal{Q}$ denote the set of queries and $\mathcal{D}$ the set of documents. For each query $q \in \mathcal{Q}$, let $\mathcal{D}_q = \{d_q^{1}, \dots, d_q^{k}\}$ denote the set of top-$k$ documents retrieved by the retriever $\mathcal{R}$. The LLM $\mathcal{M}$ can take $\mathcal{D}_q$ along with the query $q$ as input and generates an output represented as $\mathcal{O}^{rag}_q = \mathcal{M}(q, \mathcal{D}_q)$, which is evaluated using a metric $\mathcal{E}$ to calculate the accuracy of the generated response $\mathcal{O}^{rag}_q$ relative to a ground-truth $\mathcal{O}_q$ for the query $q$. In what follows, we define the notions of utility, exposure, and attribution in RAG systems.

\noindent \textbf{Utility:} In RAG, relevance does not always imply usefulness, meaning that a document may be relevant but fail to improve the accuracy of responses. In fact, since the LLM is the consumer, we observe a shift from document relevance toward document utility, which is determined by the usefulness of a document in aiding the LLM to answer the query rather than its relevance to the query. Therefore, following~\cite{kim_fairrag}, we define document utility as the marginal gain it provides to the LLM when used alongside a query. Formally, let $E_1 = \mathcal{E}(\mathcal{O}_q, \mathcal{O}^{llm}_q)$ denote the accuracy of the response $\mathcal{O}^{llm}_q$ generated in LLM-only setting, i.e., $\mathcal{O}^{llm}_q = \mathcal{M}(q)$, and let $E_2 = \mathcal{E}(\mathcal{O}_q, \mathcal{O}^{\{d_q^{i}\}}_q)$ denote the accuracy of the response $\mathcal{O}^{\{d_q^{i}\}}_q$ generated by the LLM $\mathcal{M}$ augmented with the single document $d_q^{i}$, i.e., $\mathcal{O}^{\{d_q^{i}\}}_q = \mathcal{M}(q, \{d_q^{i}\})$. The utility of document $d_q^{i}$ is defined as $u(d_q^{i}) = \max(E_2 - E_1, 0)$.

\pageenlarge 1
\noindent \textbf{Exposure:} In classical retrieval systems with human users, exposure is defined as the likelihood of a document being seen by a user which is often modeled by position-biased browsing~\cite{10.1145/3626772.3657794}, e.g., exponential decay with rank~\cite{moffat2008rank}. Such assumptions do not directly apply to machine users in RAG, since the LLM processes all top-$k$ retrieved documents. Following prior work~\cite{hsieh2024found,kim_fairrag}, we assume uniform likelihood across retrieved documents to be seen by the LLM. Formally, in RAG systems, the exposure of the document $d_q^{i} \in \mathcal{D}q$ is defined as $e(d_q^{i}) = 1$.

\noindent \textbf{Attribution:} Attribution indicates whether a retrieved document was consumed by the LLM to generate the response. Following~\cite{kim_fairrag}, we employ a NLI model, which has been shown to align well with human judgments in attribution task~\cite{honovich2022true,muller2023evaluating}. Formally, the attribution score of the document $d_q^{i} \in \mathcal{D}_q$ is defined as $a(d_q^{i}) = \text{NLI}(d_q^{i}, \mathcal{O}^{rag}_q)$, where $a(d_q^{i}) = 1$ if the document $d_q^{i}$ entails the generated response $\mathcal{O}^{rag}_q$, and $0$ otherwise, according to the NLI model.

In this work, we aim to investigate query group fairness in RAG systems. Let $\mathcal{F}=\{g_1, g_2, ..., g_f\}$ be the set of groups within a fairness category. Let $AC^{rag}(g_i)$ and $AC^{llm}(g_i)$ denote the average accuracy of responses generated for queries belonging to the group $g_i$ in the RAG and LLM-alone settings, respectively. Formally, we define two following aspects of query group fairness in RAG systems:

\noindent \textbf{\textit{Equitable Accuracy Improvements (EAI)}}: The improvements achieved by adding a retrieval component to the LLM should be equal for all groups, i.e., $\forall g_i, g_j \in \mathcal{F}, AC^{rag}(g_i) - AC^{llm}(g_i) = AC^{rag}(g_j) - AC^{llm}(g_j)$. 

\noindent \textbf{\textit{Equitable Accuracy (EA)}}: RAG systems should provide queries from all groups with the same average accuracy, i.e., $\forall g_i, g_j \in \mathcal{F}, AC^{rag}(g_i) = AC^{rag}(g_j)$.

We argue that the following key factors influence these fairness conditions:

\noindent \textbf{Group Utility:} For each group $g_i$, group utility $U(g_i)$ is defined as the average utility of documents from that group across all top-k rankings produced by the retriever for a set of queries.

\noindent \textbf{Group Exposure:} For each group $g_i$, group exposure $E(g_i)$ is defined as the average number of documents from that group presented to the LLM across all top-k rankings produced by the retriever for a set of queries.

\noindent \textbf{Group Attribution:} For each group $g_i$, group attribution $A(g_i)$ is defined as the average attribution score of documents from that group across all top-k rankings produced by the retriever for a set of queries.

Based on these definitions, our experiments examine the following directions:

\noindent \textbf{Assessing Query Group Fairness in RAG Systems:}
This direction examines whether the RAG system aligns with EAI and EA. Moreover, we compare the EA of the RAG system and LLM-only settings to determine how incorporating a retriever influences the EA of query group fairness.

\noindent \textbf{Analyzing the Role of RAG Components in Query Group Fairness:}
This direction examines how the retriever (through group exposure), generator (through group attribution) and retriever-generator interactions (through group utility) influence the query group fairness problem.

\setlength{\tabcolsep}{3pt} 
\begin{table}[tbp]
\centering
\renewcommand{\arraystretch}{0.01}
\begin{tabular}{lc}
\toprule
\multicolumn{1}{l}{\textbf{Fairness Category}} & \multicolumn{1}{c}{\textbf{Groups}} \\
\midrule
Age of the Topic (AoT) & Unk, Pre-1900s, 20th century, 21st century \\ 
\midrule
Popularity (Pop) & Low, Medium-Low, Medium-High, High \\ 
\midrule
Age of Article (AoA) & 2001–2006, 2007–2011, 2012–2016, 2017–2022 \\ 
\midrule
Alphabetical (Alp) & a–d, e–k, l–r, s–z \\ 
\bottomrule
\end{tabular}
\caption{Fairness categories and their groups.}
\label{tab:collection_1}
\end{table}
\vspace{-0.2cm}

\section{Experimental Methodology} \label{experimental_methodology}
In this work, we aim to answer the following three research questions:

\noindent \textbf{RQ1:} Do RAG systems achieve fairness in terms of EAI and EA?

\noindent \textbf{RQ2:} How does adding a retriever to the LLM affect EA?

\noindent \textbf{RQ3:} Does group exposure, attribution and utility influence unfairness in terms of EAI and EA?\\

\pageenlarge 1
\noindent \textbf{Test Collection:} We use the TREC 2022 Fair Ranking Track test collection~\cite{Ekstrand_2022_fair_ranking_overview}, which is well-suited for evaluating group-level fairness in RAG systems~\cite{wu2025does}, to construct three datasets that we use to evaluate our research questions. Each of our datasets contain queries and documents that are about a specific topic, i.e., within a specific WikiProject in the Fair Ranking Track collection. Moreover, each of the queries and documents in a dataset are associated with a single fairness group from each of four specific fairness categories from the Fair Ranking Track collection, namely: \textit{Age of the Topic}, \textit{Popularity}, \textit{Age of the Article}, and \textit{Alphabetical Order}. We select these fairness categories since they (i) span diverse characteristics such as temporal, lexical, and popularity, and (ii) have relatively few distinct group values (Table~\ref{tab:collection_1}). To construct datasets, we discard documents that exceed $512$ words to ensure that a greater number of documents are included within the LLM’s context window, while minimising truncation. We then discard documents lacking valid group labels for our selected fairness categories. Finally, to select the topics for our datasets, we rank WikiProjects (topics) by the number of documents that they contain and select the top three topics, namely \textit{Cities}, \textit{Geography}, and \textit{Military History}.

\setlength{\fboxsep}{2pt}
\begin{figure}[tb]
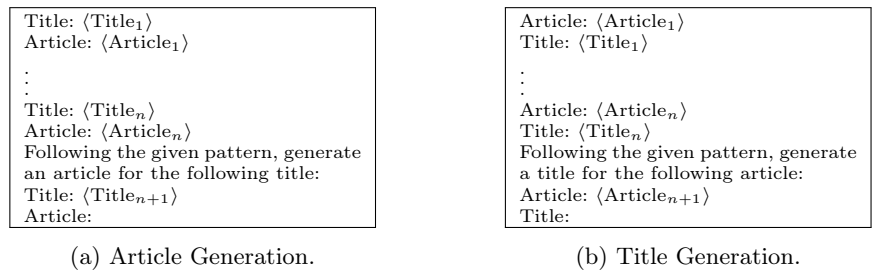

\centering
\begin{subfigure}[t]{0.46\linewidth}
\centering
\fbox{
\begin{minipage}{0.8\linewidth}
\scriptsize
Title: $\langle$Title$_1\rangle$\\
Article: $\langle$Article$_1\rangle$\\
$\vdots$\\
Title: $\langle$Title$_n\rangle$\\
Article: $\langle$Article$_n\rangle$\\
Following the given pattern, generate an article for the following title:\\
Title: $\langle$Title$_{n+1}\rangle$\\
Article:
\end{minipage}
}
\caption{Article Generation.}
\end{subfigure}
\hfill
\begin{subfigure}[t]{0.46\linewidth}
\centering
\fbox{
\begin{minipage}{0.8\linewidth}
\scriptsize
Article: $\langle$Article$_1\rangle$\\
Title: $\langle$Title$_1\rangle$\\
$\vdots$\\
Article: $\langle$Article$_n\rangle$\\
Title: $\langle$Title$_n\rangle$\\
Following the given pattern, generate a title for the following article:\\
Article: $\langle$Article$_{n+1}\rangle$\\
Title:
\end{minipage}
}
\caption{Title Generation.}
\end{subfigure}

\caption{Prompt templates used in RAG systems.}
\label{fig:prompts}
\end{figure}

\pageenlarge 1
\looseness=-1 \noindent \textbf{Tasks: }
To evaluate the query group fairness problem, we need to have queries, as well as documents, that are associated to a particular group within a fairness category. The queries from the TREC 2022 Fair Ranking Track test collection do not have such associations. Therefore, we define two contrasting tasks that use different parts of a document\footnote{Each document is associated to a specific group of a fairness category in the TREC 2022 Fair Ranking Track test collection.} as the query. In the \textit{article generation} task, a document's title is used as the query and the RAG system generates the body of the document (i.e. article), whereas in the \textit{title generation} task, the document's body is used to generate the title. For the article and title generation tasks, we use the document's article and title respectively as the ground truth for the task. In both tasks, users who seek information about underrepresented groups may receive less complete informational content, which could reinforce existing informational inequalities~\cite{jaenich2025fair}. This makes query group fairness important, since disparities in generation quality can translate into unequal access to high-quality generated content across groups. Figure~\ref{fig:prompts} illustrates the prompt templates used in this work. The two templates are structurally identical, differing only in the ordering of the article–title pairs provided to the LLM as context. Although this difference is simple, we found that maintaining the ordering of examples, as presented in Figure~\ref{fig:prompts}, resulted in markedly improved accuracy of the generated responses.

\looseness=-1 Since documents belong to multiple fairness categories, each combination of groups across fairness categories represents a distinct subset of documents. To fairly cover all subsets, we sample documents that are associated with the cartesian product of the groups within each of the fairness categories. For example, consider AoT and Pop from Table~\ref{tab:collection_1}. The cartesian product between these two fairness categories yields a set with $16$ pairs (e.g, (Unk, Low), (20th century, High)). For each member of the cartesian product, such as (Unk, Low), we select one representative document from the set of documents that simultaneously belong to Unk in AoT and Low in Pop. We do the cartesian product across all fairness categories from Table~\ref{tab:collection_1}, and select one representative document from the set of documents that simultaneously belong to each of the groups in a given member of the cartesian product. Statistics for each topic are in Table~\ref{tab:collection_2}.

\pageenlarge 1
\noindent \textbf{RAG Components:}
For the retriever component $\mathcal{R}$, we use three popular retrievers for RAG \cite{diaz2024retrieval}: BM25 (lexical) \cite{robertson1995okapi}, SPLADE (learned sparse) \cite{formal2022distillation}, and Contriever (dense bi-encoder) \cite{izacard2021unsupervised}, denoted as $\mathbf{\mathcal{B}}$, $\mathbf{\mathcal{S}}$, and $\mathbf{\mathcal{C}}$ respectively. Top-$10$ documents from each retriever are sampled as LLM context. For the generation component $\mathcal{M}$, we use two recent and widely used LLMs, namely LLaMA-3.1-8B and Gemma-2-9B, denoted as $\mathbf{\mathcal{L}}$ and $\mathbf{\mathcal{G}}$ respectively. Decoding is performed using beam search, with beam sizes of $2$ and $4$ for article generation and title generation, respectively. Following~\cite{kim_fairrag}, no sampling-based strategies are applied.

\setlength{\tabcolsep}{8pt} 
\begin{table}[tbp]
\renewcommand{\arraystretch}{0.05}
\centering
\begin{tabular}{lcccc}
\toprule
\textbf{Topic} & \textbf{\#Q} & \textbf{\#Docs} & \textbf{Avg. Body Len.} & \textbf{Avg. Title Len.} \\ 
\midrule
Cities           & 159 & 117,362 & 95.4 & 2.3 \\
Geography        & 144 & 80,273  & 65.3 & 2.2 \\
Military History & 245 & 109,169 & 251.8 & 3.2 \\ 
\bottomrule
\end{tabular}
\caption{Dataset statistics.}
\label{tab:collection_2}
\end{table}

\noindent \textbf{Evaluation Metrics:}
Let $\mathcal{Q}_{g_i} \subset \mathcal{Q}$ and $\mathcal{D}_{g_i} \subset \mathcal{D}$ denote the sets of queries and documents belonging to group $g_i$. Evaluation metrics are defined as follows:

\textbf{Query Group Accuracy:} Let $x \in \{rag, llm\}$ denote the evaluation setting, where rag represents the RAG system and llm shows the LLM-only. For each group $g_i$, the query group accuracy in setting $x$, denoted $AC^{x}(g_i)$, is defined as the average accuracy of responses generated for all queries from that group: 
\begin{equation} \label{eq:aq}
    AC^{x}(g_i) = \frac{1}{|\mathcal{Q}_{g_i}|} \sum_{q \in \mathcal{Q}_{g_i}} \mathcal{E}(\mathcal{O}_q, \mathcal{O}^{x}_q).
\end{equation}
Following~\cite{salemi2024evaluating}, we use ROUGE-L as the evaluation metric $\mathcal{E}$. We form query group accuracy vector in $x$ setting as $AC^{x} = (AC^{x}(g_i))_{g_i \in \mathcal{F}}$.

\looseness=-1\textbf{Query Group Accuracy Improvements:} Defined, for each group $g_i$, as:
\begin{equation} \label{eq:ag}
    \Delta[AC(g_i)] = AC^{rag}(g_i) - AC^{llm}(g_i).
\end{equation} 
We form query group accuracy improvements vector as $\Delta[AC] = (\Delta[AC(g_i)])_{g_i \in \mathcal{F}}$.

\textbf{Group Utility:} Measures the overall usefulness of documents from a group $g_i$ across all queries, defined as:
\begin{equation} \label{eq:u_hat}
    \hat{U}(g_i) = \frac{1}{|\mathcal{Q}|} \sum_{q \in \mathcal{Q}} \sum_{d_{q}^{j} \in D_{q}} u(d_q^{j}) \cdot \mathbf{1}_{\{d_q^{j} \in \mathcal{D}_{g_i}\}},
\end{equation}
where $\mathbf{1}_{\{condition\}} = 1$ if the condition holds. We calculate the group utility probability distribution as $U(g_i) = \frac{\hat{U}(g_i)}{\sum_{g_j \in \mathcal{F}} \hat{U}(g_j)}$, and form the utility vector $U = (U(g_i))_{g_i \in \mathcal{F}}$.

\textbf{Group Attribution:} It measures to what extent documents from a group $g_i$ are consumed by the LLM to generate the responses. Formally:
\begin{equation} \label{eq:a_hat}
    \hat{A}(g_i) = \frac{1}{|\mathcal{Q}|} \sum_{q \in \mathcal{Q}} \sum_{d_{q}^{j} \in D_{q}} a(d_q^{j}) \cdot \mathbf{1}_{\{d_q^{j} \in \mathcal{D}_{g_i}\}},
\end{equation}
where, following~\cite{kim_fairrag}, we use \textit{roberta-large-mnli}\footnote{https://huggingface.co/FacebookAI/roberta-large-mnli} model for answer attribution (i.e., $a(d_q^{i})$). Next, we calculate probability distribution of group attribution as $A(g_i) = \frac{\hat{A}(g_i)}{\sum_{g_j \in \mathcal{F}} \hat{A}(g_j)}$. We form the attribution vector as $A = (A(g_i))_{g_i \in \mathcal{F}}$.

\textbf{Group Exposure:} Measures to what extent documents from a group $g_i$ appear in the top-$k$ list presented to the LLM, defined as:
\begin{equation} \label{eq:e}
    \hat{E}(g_i) = \frac{1}{|\mathcal{Q}|} \sum_{q \in \mathcal{Q}} \sum_{j=1}^{k} e(d_q^{j}) \cdot \mathbf{1}_{\{d_q^{j} \in \mathcal{D}_{g_i}\}}.
\end{equation}
We calculate the probability distribution of group exposure as $E(g_i) = \frac{\hat{E}(g_i)}{\sum_{g_j \in \mathcal{F}} \hat{E}(g_j)}$, and form the exposure vector as $E = (E(g_i))_{g_i \in \mathcal{F}}$.

\setlength{\belowcaptionskip}{0pt} 
\setlength{\tabcolsep}{2pt} 
\begin{table*}[tbp]
\renewcommand{\arraystretch}{0.85}
\small
\centering
\begin{minipage}{0.41\linewidth}
\centering
\begin{tabular}{@{}ccc|c|c|c@{}}
\toprule
\multirow{2}{*}{Task} & 
\multirow{2}{*}{\rotatebox{90}{LLM}} & 
\multicolumn{4}{c}{Retriever} \\
\cmidrule(lr){3-6}
 & &  N & $\mathcal{B}$ & $\mathcal{S}$ & $\mathcal{C}$ \\
\toprule
\multirow{2}{*}{\shortstack{Article\\Generation}} & $\mathcal{L}$ & 21.8 & 28.4 & 28.9 & 29.0\\
                     & $\mathcal{G}$ & 23.5 & 28.6 & \textbf{29.3} & \textbf{29.3}\\
\midrule                     
\multirow{2}{*}{\shortstack{Title\\Generation}} & $\mathcal{L}$ & 13.1 & 20.5 & 20.2 & 20.5\\
                     &  $\mathcal{G}$ & 49.4 & 74.3 & \textbf{77.7} & 74.8\\
\bottomrule
\end{tabular}
\subcaption{Cities}
\end{minipage}%
\hspace{0.00001\linewidth} 
\begin{minipage}{0.32\linewidth}
\centering
\begin{tabular}{@{}c|c|c|c@{}}
\toprule
\multicolumn{4}{c}{Retriever} \\
\cmidrule(lr){1-4}
N & $\mathcal{B}$ & $\mathcal{S}$ & $\mathcal{C}$ \\
\toprule
22.2 & 32.1 & \textbf{33.1} & 32.4\\
23.2 & 28.9 & 32.1 & 31.0\\
\midrule
11.6 & 26.4 & 26.3 & 27.1\\
54.2 & \textbf{90.2} & 88.3 & 88.3\\
\bottomrule
\end{tabular}
\subcaption{Geography}
\end{minipage}
\begin{minipage}{0.25\linewidth}
\centering
\begin{tabular}{@{}c|c|c|c@{}}
\toprule
\multicolumn{4}{c}{Retriever} \\
\cmidrule(lr){1-4}
N & $\mathcal{B}$ & $\mathcal{C}$ & $\mathcal{S}$ \\
\toprule
20.0 & 23.2 & \textbf{24.8} & 23.7\\
19.9 & 20.4 & 22.0 & 21.5\\
\midrule
11.5 & 29.4 & 28.9 & 28.6\\
54.6 & \textbf{84.1} & 82.2 & 82.0\\
\bottomrule
\end{tabular}
\subcaption{Military History}
\end{minipage}
\caption{Average ROUGE-L scores in the RAG and LLM-only (N) settings. The best LLM-retriever combination is highlighted bold for each task and dataset.}
\label{tab:base_rouge}
\end{table*}

\pageenlarge 1
\section{Experimental Results} \label{experimental_results}
In our experiments, as can be seen from Table~\ref{tab:base_rouge}, for both the article generation and title generation tasks the RAG system improves the average ROUGE-L scores of responses across all queries, compared to the LLM-only setting, when any of the retrievers (BM25 $\mathbf{\mathcal{B}}$, SPLADE $\mathbf{\mathcal{S}}$ or Contriever $\mathbf{\mathcal{C}}$) are deployed. Moreover, Gemma-2-9B (LLM $\mathbf{\mathcal{G}}$) in the majority of experiments outperforms LLaMA-3.1-8B (LLM $\mathbf{\mathcal{L}}$) in both the RAG and LLM-only settings.

\setlength{\belowcaptionskip}{0pt} 
\begin{figure}[tbp]
    \centering
    \includegraphics[width=1\linewidth]{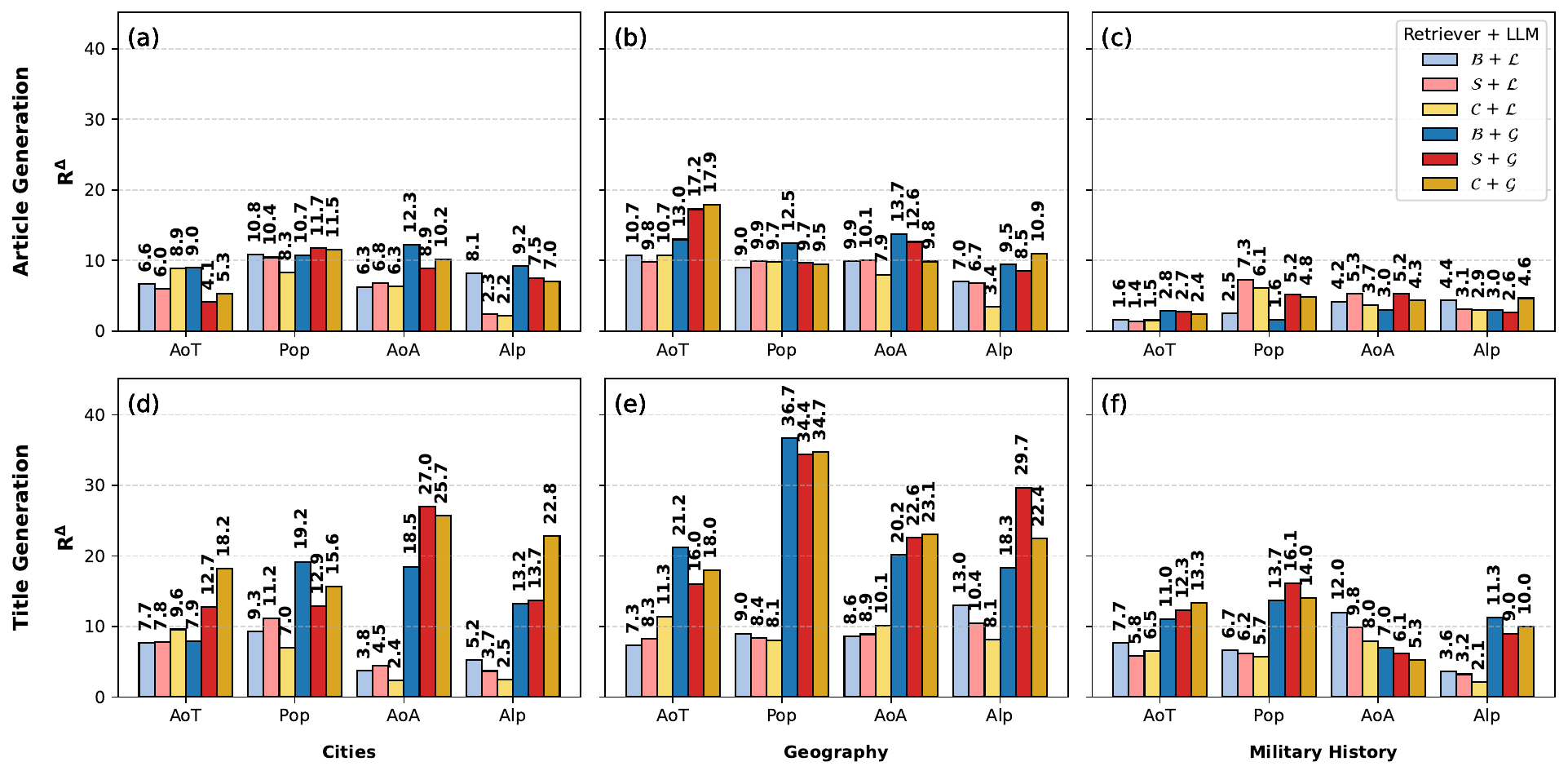}
    \caption{Range of query group accuracy gains, i.e., $R^{\Delta}$, for all fairness categories across all topics, tasks, retrievers and LLMs.}
    \label{fig:EAG}
\end{figure}

\noindent \textbf{Assessing Query Group Fairness in RAG Systems (RQ1 \& RQ2):} To answer RQ1, we investigate whether the RAG systems result in \textit{Equitable Accuracy Improvements} (EAI) and \textit{Equitable Accuracy} (EA). To assess EAI, for each topic–task–retriever–LLM combination, we calculate the range of query group accuracy improvements across all groups within a fairness category, i.e., $R^{\Delta} = \max_{g_i \in \mathcal{F}} \Delta[AC(g_i)] \,-\, \min_{g_i \in \mathcal{F}} \Delta[AC(g_i)]$. A zero range corresponds to complete fairness, while higher ranges signify greater unfairness in EAI. Figure~\ref{fig:EAG} presents these ranges for each fairness category. The highest unfairness (range = $36.7$) occurs in Figure~\ref{fig:EAG}e for the Pop fairness category on the Geography topic when we use LLM $\mathcal{G}$ and retriever $\mathcal{B}$ for title generation. In contrast, the lowest unfairness (range = $1.4$) is in Figure~\ref{fig:EAG}c for AoT fairness category on Military History topic when we use LLM $\mathcal{L}$ and retriever $\mathcal{S}$ for article generation. Across tasks, title generation exhibits larger ranges, sometimes exceeding $30$ $\Delta[Acc(g_i)]$ points (e.g., Geography topic, Pop fairness category in Figure~\ref{fig:EAG}e) whereas, article generation generally yields smaller ranges, with several cases having values below $10$ $\Delta[Acc(g_i)]$ points (e.g., in Military History topic, AoT fairness category in Figure~\ref{fig:EAG}c). Thus, unfairness under the EAI aspect is more pronounced in the title generation task, i.e., for shorter generated content, than in the article generation task. This indicates that shorter generated outputs are more sensitive to biases introduced by RAG, potentially because there is less contextual space for the model to incorporate diverse group information. Across fairness categories, Pop and AoA often show the largest ranges in each topic. Across topics, clear differences emerge: Geography often shows the highest ranges (Figures~\ref{fig:EAG}b, e), particularly in Pop and AoA fairness categories, while Military History exhibits the smallest ranges for most fairness categories (Figures~\ref{fig:EAG}c, f). Across generation components, LLM $\mathcal{G}$ is less fair than LLM $\mathcal{L}$ in most cases. For example, for approaches that use the same retriever in Figure~\ref{fig:EAG}(f), i.e., bars that are different shades of the same colour (dark for LLM $\mathcal{G}$ and light for LLM $\mathcal{L}$), LLM $\mathcal{L}$ is less fair than LLM $\mathcal{G}$ for all fairness categories except AoA. This suggests that a LLM's architecture or training data can affect the LLM's sensitivity to group-level bias within the retrieved documents, and therefore fairness outcomes are not solely determined by the retriever.

\begin{figure}[tbp]
    \centering
    \includegraphics[width=1\linewidth]{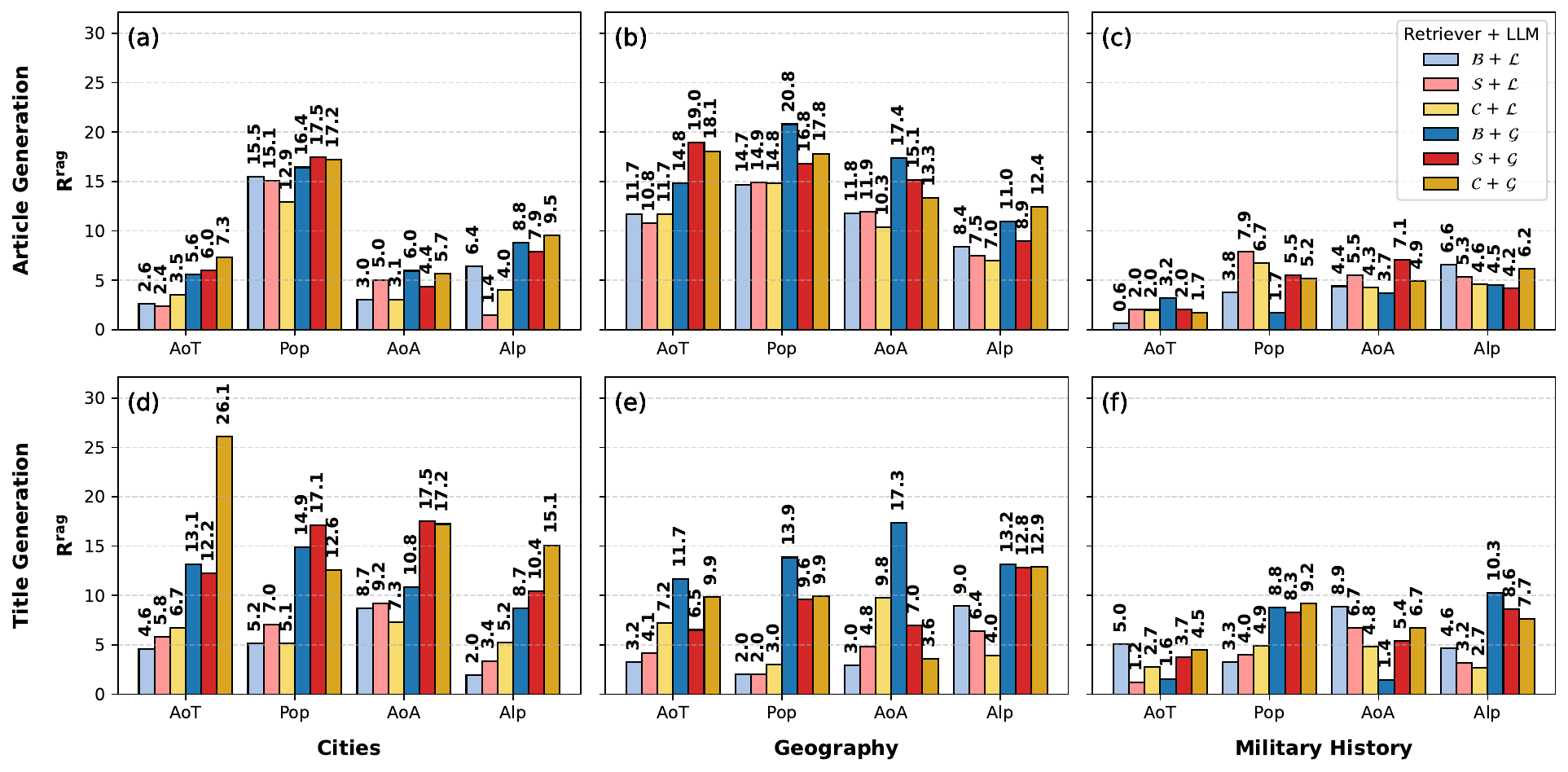}
    \caption{Range of query group accuracy in a RAG setting, i.e., $R^{rag}$, for all fairness categories across all topics, tasks, retrievers and LLMs.}
    \label{fig:ea_rag}
\end{figure}

\pageenlarge 1
\looseness=-1 To assess EA, for each topic–task–retriever–LLM combination, we compute the range of query group accuracy in RAG setting across all groups within a fairness category, i.e., $R^{rag} = \max_{g_i \in \mathcal{F}} AC^{rag}(g_i)\,-\,\min_{g_i \in \mathcal{F}} AC^{rag}(g_i)$. A zero range corresponds to complete fairness, while higher ranges signify more unfairness. Figure~\ref{fig:ea_rag} presents these ranges. According to Figure~\ref{fig:ea_rag}c, the lowest unfairness (range = $0.6$) occurs in the Military History topic for the article generation task using retriever $\mathcal{B}$ and LLM $\mathcal{L}$, on AoT fairness category. In contrast, the highest unfairness (range = $26.1$) is in Figure~\ref{fig:ea_rag}d in Cities topic for title generation on the AoT fairness category with retriever $\mathcal{C}$ and LLM $\mathcal{G}$. Across fairness categories, Pop exhibits the highest unfairness in the article generation task, with ranges of $17.5$, $20.8$, $7.9$ in Figures~\ref{fig:ea_rag}a, b and c, respectively. However, there is not any uniform pattern across fairness categories for the title generation task (i.e., Figures~\ref{fig:ea_rag}d, e, and f). From a generator perspective, we compare the bars that are different shades of the same colour in each task-topic-fairness category combination. Interestingly, LLM $\mathcal{G}$ is less fair than LLM $\mathcal{L}$ in most combinations. For example, Figure~\ref{fig:ea_rag}(a) shows that LLM $\mathcal{L}$ is less fair than LLM $\mathcal{G}$ only in AoA fairness category and retriever $\mathcal{S}$. Across topics, Military History demonstrates the lowest degree of unfairness in both tasks (see Figures~\ref{fig:ea_rag}c, f). Moreover, according to Figure~\ref{fig:ea_rag}b, article generation on Geography topic appears most challenging, as all retriever-LLM combinations show relatively high unfairness across all fairness categories, as opposed to other sub-figures which show relatively low unfairness in some cases. In conclusion, RAG systems fail to meet EAI and EA in our experiments.

\pageenlarge 1
To answer RQ2, we investigate whether the use of a retriever in RAG mitigates or amplifies EA compared to the LLM-only setting. Table~\ref{tab:RQ2} compares the range of query group accuracy for the RAG ($R^{rag}$) and LLM-only ($R^{llm}$) settings in article and title generation tasks. We calculate the range of query group accuracy in the LLM-only setting as $R^{llm} = \max_{g_i \in \mathcal{F}} AC^{llm}(g_i) \,-\, \min_{g_i \in \mathcal{F}} AC^{llm}(g_i)$. A zero range corresponds to complete fairness, while higher ranges signify more unfairness. From Table~\ref{tab:RQ2}c, in terms of topics, we see that Military History shows the lowest unfairness in both the RAG and LLM-only (denoted as N) settings, and in both tasks. Across fairness categories, Pop emerges as the most challenging category for EA, consistently exhibiting high levels of unfairness in both settings and across both tasks for all topics. In the LLM-only setting, for both tasks, we observe unfairness in all fairness categories and across all topics, in terms of the EA fairness aspect. Overall, in the LLM-only setting, LLM $\mathcal{G}$ results in more unfairness than LLM $\mathcal{L}$. Additionally, title generation is often more unfair than article generation, indicating that shorter outputs are more sensitive to biases from the LLM. Finally, when comparing RAG and LLM-only settings, we mostly observe an increase in unfairness when adding any retriever to the LLM, with some exceptions. For example, in Table\ref{tab:RQ2}c, adding a retriever to LLM $\mathcal{G}$ for the title generation task decreases the unfairness from $10.8$ in the LLM-only setting to $1.6$. Therefore, while retrieval can occasionally reduce disparities by providing more balanced content, in many cases it can amplify existing biases.

\setlength{\belowcaptionskip}{0pt} 
\setlength{\tabcolsep}{2pt} 
\begin{table*}[tbp]
\renewcommand{\arraystretch}{0.85}
\centering
\begin{minipage}{0.4\linewidth}
\centering
\begin{tabular}{@{}cccc|c|c|c@{}}
\toprule
\multirow{2}{*}{\rotatebox{90}{Task}} & 
\multirow{2}{*}{\rotatebox{90}{Cat.}} & 
\multirow{2}{*}{\rotatebox{90}{LLM}} & 
\multicolumn{4}{c}{Retriever} \\
\cmidrule(lr){4-7}
 & & & N & $\mathcal{B}$ & $\mathcal{C}$ & $\mathcal{S}$ \\
\midrule
\multirow{8}{*}{\rotatebox{90}{Article Generation}} & \multirow{2}{*}{\rotatebox{90}{AoT}} & $\mathcal{L}$ & 5.7 & 2.6 & 2.4 & 3.5\\
                     & & $\mathcal{G}$ & 6.7 & 5.6 & 6.0 & 7.3\\
\cmidrule(lr){2-7}
& \multirow{2}{*}{\rotatebox{90}{Pop}} & $\mathcal{L}$ & 4.7 & 15.5 & 15.1 & 12.9\\
                     & & $\mathcal{G}$ & 6.9 & 16.4 & 17.5 & 17.2\\
\cmidrule(lr){2-7}
& \multirow{2}{*}{\rotatebox{90}{AoA}} & $\mathcal{L}$ & 3.4 & 3.0 & 5.0 & 3.1\\
                     & & $\mathcal{G}$ & 6.3 & 6.0 & 4.4 & 5.7\\
\cmidrule(lr){2-7}
& \multirow{2}{*}{\rotatebox{90}{Alp}} & $\mathcal{L}$ & 2.6 & 6.4 & 1.4 & 4.0\\
                     & & $\mathcal{G}$ & 3.0 & 8.8 & 7.9 & 9.5\\
\Xhline{.9pt}
\multirow{8}{*}{\rotatebox{90}{Title Generation}} & \multirow{2}{*}{\rotatebox{90}{AoT}} & $\mathcal{L}$ & 4.3 & 4.6 & 5.8 & 6.7\\
                     & & $\mathcal{G}$ & 13.3 & 13.1 & 13.2 & 26.1\\
\cmidrule(lr){2-7}
& \multirow{2}{*}{\rotatebox{90}{Pop}} & $\mathcal{L}$ & 4.1 & 5.2 & 7.0 & 5.1\\
                     & & $\mathcal{G}$ & 14.0 & 14.9 & 17.1 & 12.6\\
\cmidrule(lr){2-7}
& \multirow{2}{*}{\rotatebox{90}{AoA}} & $\mathcal{L}$ & 5.0 & 8.7 & 9.2 & 7.3\\
                     & & $\mathcal{G}$ & 13.0 & 10.8 & 17.5 & 17.2\\
\cmidrule(lr){2-7}
& \multirow{2}{*}{\rotatebox{90}{Alp}} & $\mathcal{L}$ & 7.1 & 2.0 & 3.4 & 5.2\\
                     & & $\mathcal{G}$ & 8.8 & 8.7 & 10.4 & 15.1\\
\bottomrule
\end{tabular}
\subcaption{Cities}
\end{minipage}%
\begin{minipage}{0.25\linewidth}
\centering
\begin{tabular}{@{}c|c|c|c@{}}
\toprule
\multicolumn{4}{c}{Retriever} \\
\cmidrule(lr){1-4}
N & $\mathcal{B}$ & $\mathcal{C}$ & $\mathcal{S}$ \\
\midrule
3.7 & 11.7 & 10.8 & 11.7\\
6 & 14.8 & 19.0 & 18.1\\
\cmidrule(lr){1-4}
5.7 & 14.7 & 14.9 & 14.8\\
8.3 & 20.8 & 16.8 & 17.8\\
\cmidrule(lr){1-4}
3.6 & 11.8 & 11.9 & 10.3\\
5.0 & 17.4 & 15.1 & 13.3\\
\cmidrule(lr){1-4}
3.6 & 8.4 & 7.5 & 7.0\\
4.0 & 11.0 & 8.9 & 12.4\\
\Xhline{.9pt}
5.1 & 3.2 & 4.1 & 7.2\\
9.5 & 11.7 & 6.5 & 9.9\\
\cmidrule(lr){1-4}
7.0 & 2.0 & 2.0 & 3.0\\
24.8 & 13.9 & 9.6 & 9.9\\
\cmidrule(lr){1-4}
7.1 & 3.0 & 4.8 & 9.8\\
25.6 & 17.3 & 7.0 & 3.6\\
\cmidrule(lr){1-4}
4.1 & 9.0 & 6.4 & 4.0\\
20.7 & 13.2 & 12.8 & 12.9\\
\bottomrule
\end{tabular}
\subcaption{Geography}
\end{minipage}
\begin{minipage}{0.25\linewidth}
\centering
\begin{tabular}{@{}c|c|c|c@{}}
\toprule
\multicolumn{4}{c}{Retriever} \\
\cmidrule(lr){1-4}
N & $\mathcal{B}$ & $\mathcal{C}$ & $\mathcal{S}$ \\
\midrule
1.7 & 0.6 & 2.0 & 2.0\\
1.0 & 3.0 & 2.0 & 1.7\\
\cmidrule(lr){1-4}
1.3 & 3.8 & 7.9 & 6.7\\
0.6 & 1.7 & 5.5 & 5.2\\
\cmidrule(lr){1-4}
1.8 & 4.4 & 5.5 & 4.3\\
2.0 & 3.7 & 7.1 & 4.9\\
\cmidrule(lr){1-4}
2.2 & 6.6 & 5.3 & 4.6 \\
2.1 & 4.5 & 4.2 & 6.2\\
\Xhline{.9pt}
5.1 & 5.0 & 1.2 & 2.7\\
10.8 & 1.6 & 3.7 & 4.5\\
\cmidrule(lr){1-4}
5.2 & 3.3 & 4.0 & 4.9\\
15.5 & 8.8 & 8.3 & 9.2\\
\cmidrule(lr){1-4}
6.5 & 8.9 & 6.7 & 4.8\\
8.0 & 1.4 & 5.4 & 6.7\\
\cmidrule(lr){1-4}
4.0 & 4.6 & 3.2 & 2.7 \\
7.0 & 10.3 & 8.6 & 7.7\\
\bottomrule
\end{tabular}
\subcaption{Military History}
\end{minipage}
\caption{Comparison of the range of query group accuracy in the RAG and LLM-only (indicated as N) settings, i.e., $R^{rag}$ vs $R^{llm}$.}
\label{tab:RQ2}
\end{table*}

In conclusion, RAG systems fail to meet EAI and EA for query group fairness and, in most cases, amplify EA unfairness compared to the LLM-only setting. We investigate the potential causes of this unfairness in RQ3.

\setlength{\belowcaptionskip}{0pt} 
\begin{figure}[tbp]
    \centering
    \includegraphics[width=0.9\linewidth]{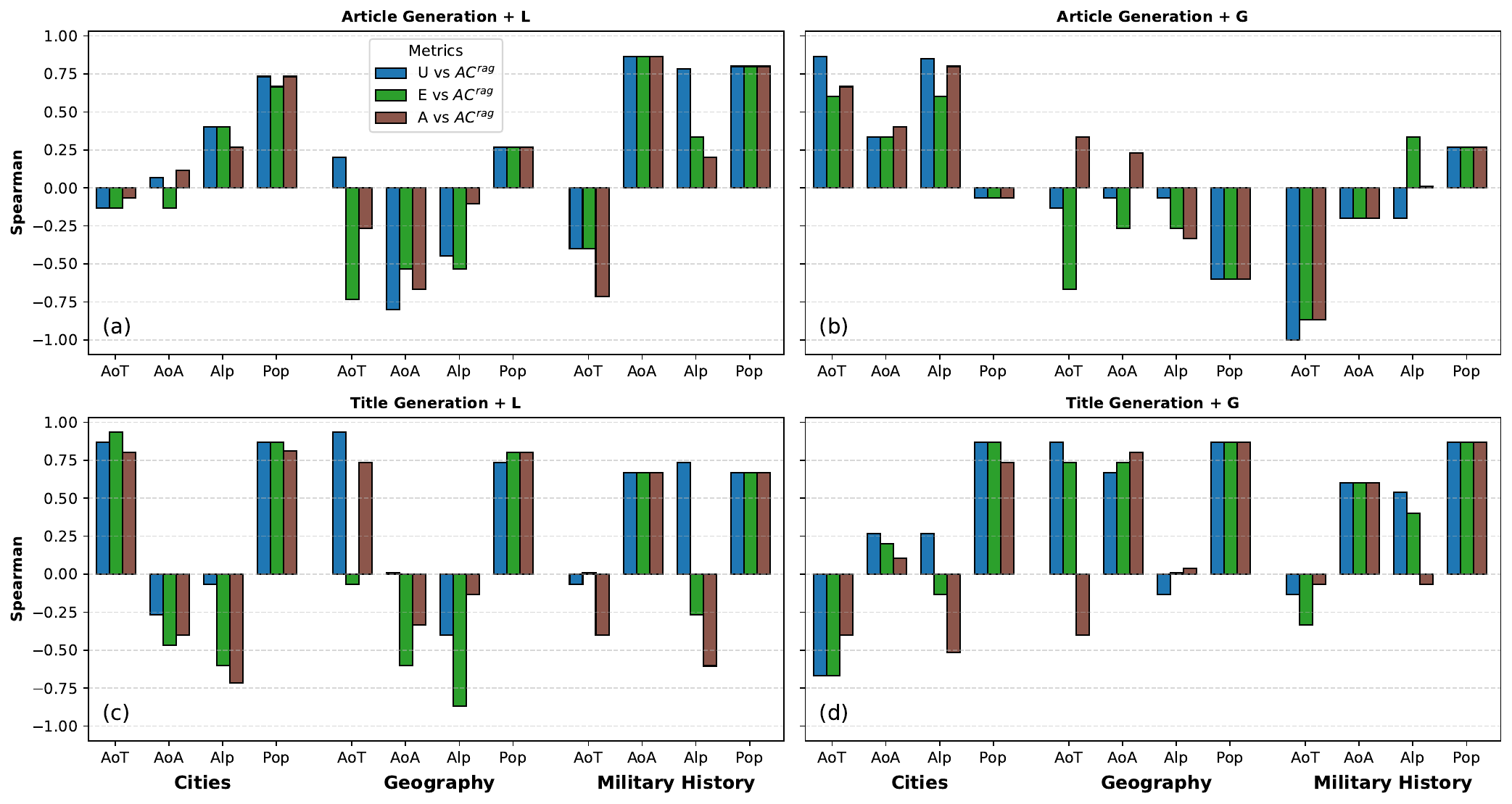}
    \caption{Spearman correlation between query group accuracy in RAG setting (i.e., $AC^{rag}(g_i)$) and corresponding group utility, exposure and attribution for all fairness categories across topics, task, and LLMs.}
    \label{fig:rq3_ea}
\end{figure}

\begin{figure}[tbp]
    \centering
    \includegraphics[width=0.9\linewidth]{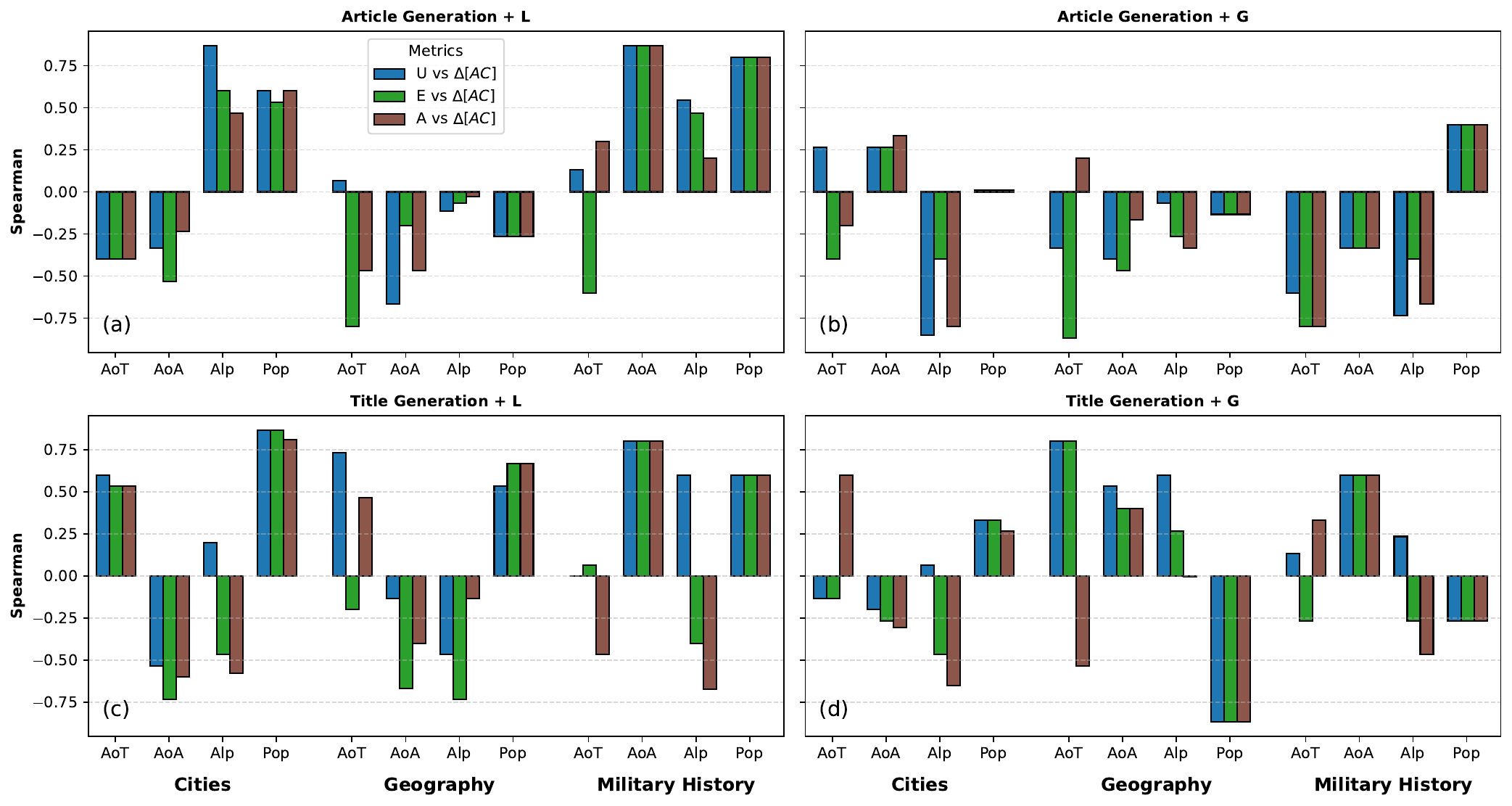}
    \caption{Spearman correlation between query group accuracy gain (i.e., $\Delta(AC(g_i))$) and corresponding group utility, exposure and attribution for all fairness categories across topics, tasks, and LLMs.}
    \label{fig:rq3_eag}
\end{figure}

\pageenlarge 1
\looseness=-1\noindent \textbf{Analyzing the Role of RAG Components in Query Group Fairness (RQ3):} To answer RQ3, we examine how the retriever (captured by group exposure), the generator (captured by group attribution), and the retriever-generator interaction (captured by group utility) relate to EAI and EA. For each topic-task-LLM combination, we compute Spearman correlations between (i) $U$ and $AC^{rag}/\Delta[AC]$, (ii) $E$ and $AC^{rag}/\Delta[AC]$, and (iii) $A$ and $AC^{rag}/\Delta[AC]$, separately for each retriever. Then, correlations are averaged across retrievers to provide aggregated measures of how query group accuracy and query group accuracy improvements are influenced by the corresponding group utility, exposure and attribution. Figures~\ref{fig:rq3_ea} and \ref{fig:rq3_eag} present these averaged correlations for all fairness categories in each topic-task-LLM combination. According to the figures, when comparing group utility, exposure, and attribution within each fairness category, we observe that their correlations with query group accuracy or query group accuracy improvements typically align in direction which can be positive or negative. For instance, in Figure~\ref{fig:rq3_ea}b, this pattern holds consistently, except in three out of $12$ cases: Geography topic under the AoT and AoA categories, and Military History under the Alp category. We sometimes observe positive correlations in both figures, for instance, in AoA and Pop fairness categories within the Military History topic in Figure~\ref{fig:rq3_eag}b, and AoT and Pop categories within the Cities topic in Figure~\ref{fig:rq3_ea}c. Such a correlation indicates that the RAG system tends to prioritise or attribute credit to groups whose documents genuinely enhance answer quality. In these cases, documents from such groups are not only frequently retrieved (i.e., high utility and exposure) but also effectively consumed by the LLM (i.e., high attribution). However, negative correlations are also observed in both figures, for example, in AoT fairness category within Cities topic in Figure~\ref{fig:rq3_ea}d, and in AoT and AoA fairness categories within Cities topic in Figure~\ref{fig:rq3_eag}a. Such a correlation indicates that higher query group accuracy or query group accuracy improvements for a particular group corresponds to lower utility, exposure, and attribution for that group. This pattern can be explained by the room-for-improvement effect: groups with lower query group accuracy in the LLM-only setting (i.e., $AC^{llm}(g_i)$) have greater potential for improvement in the RAG setting, even when their exposure, utility, or attribution scores are relatively small. Overall, these findings highlight that group utility, exposure, and attribution can all influence query group fairness issue in RAG systems.

\pageenlarge 1
\section{Conclusion} \label{conclusion}
In this work, we introduced the query group fairness problem in RAG systems. We investigated two aspects  (i) EAI: ensuring equitable average accuracy in responses across queries for different groups, and (ii) EA: ensuring equitable average accuracy improvements across queries from different groups. Using three datasets built from the TREC 2022 Fair Ranking Track, we evaluated the problem on two tasks and four fairness categories. Our findings showed that RAG systems suffer from query group fairness issues. Moreover, we analysed the effect of group exposure, attribution and utility on query group fairness. Our results showed that, in most cases, whenever useful documents from a particular group are retrieved (high utility and exposure) and consumed (high attribution) by RAG systems, the corresponding query group accuracy and accuracy improvements are also high. This work demonstrates the importance of group utility, exposure and attribution in the development of fair RAG systems.

\section*{Acknowledgments}
This work was supported by the Engineering and Physical Sciences Research Council grant number EP/Y009800/1, through funding from Responsible Ai UK (KP0011).

\section*{Disclosure of Interests}
The authors have no competing interests to declare that are relevant to the content of this article.

%
%
%
%

\bibliographystyle{splncs04} 
\bibliography{references}






\end{document}